# Metamaterial slab-based super-absorbers and perfect nanodetectors for single dipole sources

**Guang-Yu Guo,[1,2,*] Vasily Klimov,[3,4] Shulin Sun,[1,5] and Wei-Jin Zheng[1]**

[1]*Department of Physics, National Taiwan University, Taipei 10617, Taiwan*
[2]*Graduate Institute of Applied Physics, National Chengchi University, Taipei 11605, Taiwan*
[3]*P. N. Lebedev Physical Institute, Russian Academy of Sciences,119991 Moscow, Russia*
[4]*vklimov@sci.lebedev.ru*
[5]*National Center of Theoretical Sciences, Physics Division, National Taiwan University, Taipei 10617, Taiwan*

*\*gyguo@phys.ntu.edu.tw*

**Abstract:** We propose to use double negative (DNG) metamaterial slabs to build effective super-absorbers and perfect nanodetectors for single divergent sources. We demonstrate by numerical simulations that an absorbing nanoparticle properly placed inside a DNG slab back-covered with a perfect electric conductor or perfect magnetic conductor mirror can absorb up to 100% radiation energy of a single dipole source placed outside the slab. Furthermore, wealso show that even the simple DNG slab without any absorbing nanoparticle could be used as a perfect absorber for both plane and divergent beams. The proposed systems may focus the radiation in nanoscale and thus have applications in optical nanodevices for a variety of different purposes.

## References and links

# 1. Introduction

A novel concept of coherent perfect absorbers (CPAs) [1,2] has recently been proposed and also demonstrated experimentally by the example of asymetric Fabry-Perot cavity with small losses which was symmetrically irradiated by 2 coherent plane waves. All the energy of incoming plane waves is fully absorbed by the dielectric slab, and hence the radiation from the system disappears for certain parameters. More complicated examples of the CPA with plane waves were considered in [3-5].

Despite generic nature of this concept, its application to more sophisticated geometries is a nontrivial task. Since the control of radiation of atoms and molecules with nanoparticles and metamaterialsis an important topic today, it is desirable to have an effective nanoabsorber or nanodetector of electromagnetic fields from a single atom or molecule. Therefore, based on both analytical dipole model analyses and numerical simulations, we have recently proposed a coherent perfect nanoabsorber (CPNA) for 2 divergent beams [6] using a slab made of double negative (DNG) metamaterial with negative refraction [7]. This device would allow absorbing all the energy from the 2 point dipole sources emitted in the direction towards the negative refraction slab by a lossy nanoparticle placed inside it. In [6], the CPNA system is symmetric and works with two coherent dipole sources with the same orientation and placed symmetrically on both sides of the slab. Therefore, for practical uses of this device to focus near-field light, one may encounter the difficulties in accurately putting the two radiating dipoles at the symmetric positions and also making them radiate coherently.

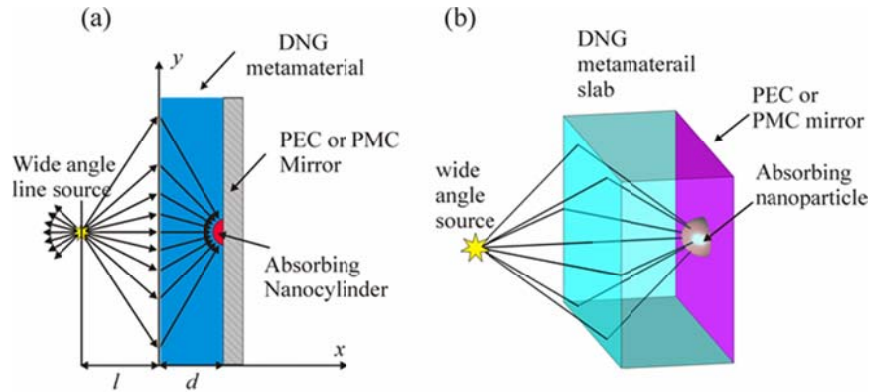

Fig. 1. Schematic diagram of a perfect nanoabsorber for a divergent beam [(a): 2D case and (b): 3D case]. The source is situated at one side of the double negative (DNG) slab. The distance $l$ between the source and the surface is equal to its thickness $d$. The other side of the DNG slab is covered with a PEC or PMC mirror with an absorbing nanoparticle on it.

In this paper, we propose another system which deals only with one wide-angle point source to overcome the difficulties related to the symmetry and coherence. The geometry of our device is shown in Fig. 1.It follows from Fig. 1 that the wide angle source is situated at one side of the DNG slab and the distance between the source and DNG slab is equal to its thickness. The other side of the DNG slab is covered with a perfect electric conductor (PEC) or perfect magnetic conductor (PMC) mirror with an absorbing nanoparticle on it. Here, by PEC we mean a hypothetic substance with permittivity $\varepsilon \to \infty$ while permeability $\mu \neq 0$. Many metals are in fact very close to this definition. PMC is also a hypothetic substance with permeability $\mu \to \infty$ but permittivity $\varepsilon$ remains finite. Nevertheless, PMC is more exotic than PEC. However, it is possible to manufacture materials whose properties are close to that of PMC, thanks to recent technology developments [8].

We will consider two different orientations of the source dipole momentum. In general, the geometry shown in Fig. 1 cannot be reduced to symmetric geometry of the CPNA in [6]. Moreover, as we will see later, such a geometry allows to absorb about 100% of light emitted in all directions for some combinations of the dipole orientation and the mirror material. For simplicity, we will restrict our simulations to the 2D case, and the generalization to the 3D case is straightforward. Since an exact solution for the lossless 3D case which can be described within a modified image model has also been reported [9], we believe that our conclusions drawn from our 2D modeling would also be valid for the 3D systems with small losses and an absorbing particle instead of a point energy sink.

The rest of the paper is organized as follows. In section 2, we will consider different combinations of the dipole orientations and mirror materials, and show that there are parameters regions where more than half of the emitted energy is absorbed by the nanoparticle and the system can be considered as a perfect nanodetector. In section 3, we will consider the system shown in Fig. 1(a) without the nanocylinder, and demonstrate that the DNG slab alone could absorb almost 100% of the emitted energy. Finally, a summary of this work will be given in section 4.

**2. Perfect nanoabsorbers and nanodetectors**

As for the CPNAs [6], the concepts of the present perfect nanodetectors and nanoabsorbers are based on the exact solutions of the Maxwell equations for *lossless* DNG media found earlier for the 3D [9] and 2D [10] cases. Due to the unusual properties of negative refraction ($n = -1$) media, these solutions contain both usual sources of fields and also singular sinks of energy. In [9-10] it was shown that all sources and sinks of energy are situated at the points corresponding to a picture of image charges. Therefore, we will use this simple picture of image charges to understand the results of our simulations throughout this paper. However, it should be pointed out that the conditions of refractive index $n = -1$ and absence of losses are of crucial importance because only in this case a picture of image charges is valid. Physically, this occurs because the absolute values of refractive index for vacuum and the negative refraction medium are equal. Nonetheless, our recent numerical simulations for the CPNAs [6] confirmed the validity of this image charge picture also for real systems made of DNG metamaterials with negative refraction and losses.

To investigate the feasibility and also understand the operation principle of the above proposed devices, we have performed electromagnetic simulations of the system shown in Fig. 1(a), using the finite elements method (FEM) within the COMSOL Multiphysics® software. For simplicity were strict ourselves to a 2D DNG slab with small losses $\varepsilon_s = \mu_s = -1 + i\varepsilon_s''$ (region 2) placed in vacuum with $\varepsilon_0 = 1$ and $\mu_0 = 1$ (region 1). We use a nanocylinder with parameters $\varepsilon_c = 1 + i\varepsilon_c''$ and $\mu_c = 1$, as the absorbing (or detecting) element.

*2.1 Source dipole parallel to the metamaterial slab*

Let us first consider the case of the emitting electric dipole being parallel to the surface of the metamaterial slab. In this case, the dipole nanowire has a current density given by

$$j_y = -i\omega p_0 \delta(y)\delta(x+l) \qquad (1)$$

where $p_0$ is the dipole moment of the nanowire per unit length, and the $e^{-i\omega t}$ dependence is assumed throughout. That is, we are considering a 2D geometry with only nonzero $H_z$ component of the magnetic field. The (normalized) power absorbed by the nanoparticle ($W_c/W_0$), the slab ($W_s/W_0$) and by the whole system (i.e., the nanoparticle plus the slab) ($W_{c+s}/W_0$) as well as the (normalized) power radiated from the source dipole ($W_p/W_0$) for the systems with either a PMC or PEC mirror, are displayed in Fig. 2. Here $W_0$ is the power emitted by the dipole per unit length in free space

$$W_{rad} = \frac{\omega}{2} k_0^2 |p_0|^2 \frac{\pi}{2}. \qquad (2)$$

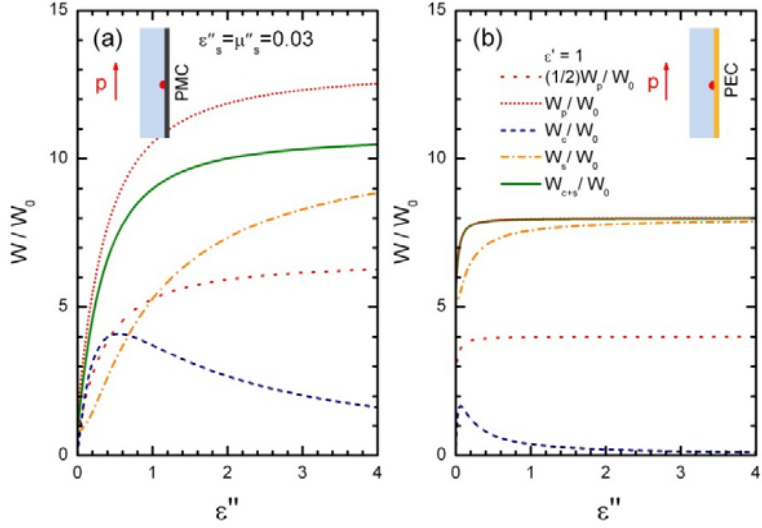

Fig. 2. Powers absorbed by the nanoparticle ($W_c/W_0$), the slab ($W_s/W_0$) and the system (nanoparticle plus slab) ($W_{c+s}/W_0$) normalized to the radiating power of one dipole in free space ($W_0$) as a function of the imaginary part of the cylinder permitivity for the system with either a PMC mirror (a) or a PEC mirror (b) at the back of the metamaterial slab. The radiating electric dipole is parallel to the surface of the DNG slab. The total ($W_p/W_0$) and half [($W_p/2)/W_0$] normalized radiating powers of the dipole are also plotted for comparison. The radius of nanoparticle ($r_c$) is 50 nm, and its optical parameters are $\varepsilon_c = 1 + i\varepsilon''$ and $\mu_c = 1$, $\varepsilon_s = \mu_s = -1 + 0.03i$, $\lambda = 3$ μm, $d = 250$ nm and $l = 250$ nm.

First of all, one can see that total emitted power is substantially greater than in the case of free space. This is a manifestation of Purcell effect [11,12]. From the physical point of view, this enhancement is due to additional channel (surface plasmon wave) into which excited atom can decay. Another important feature of Fig. 2(a) (the PMC mirror case) is that the power spectra of the absorbing nanoparticle and the (half) emitting dipole are identical to that of the 2 symmetric source system considered before [6], whereas in contrast, the PEC mirror system exhibits a different behavior from the former two systems. In particular, the energy absorbed by the nanoparticle in the PMC system in the region of $0.1 < \varepsilon'' < 0.5$ is greater than half of the energy radiated by a single source [($W_p/2)/W_0$], implying that the nanoparticle can be considered as a perfect nanoabsorber, as proposed before [6]. Alternatively, the system can also be regarded as a perfect nanodetector if the nanocylinder is a detecting element. In contrast, the power absorbed by the nanocylinder in the PEC system[Fig. 2(b)] is significantly smaller than half of the energy radiated by a single source [($W_p/2)/W_0$] in the entire region of the imaginary part of the cylinder permittivity considered here. In other words, the PEC system does not exhibit perfect nanoabsorptionn or nanodetection behavior. On the other hand, in the PEC system, all emitted power is absorbed by the whole system for almost all values of $\varepsilon''$. This remarkable fact will be further discussed in section 3.

These different behaviors of the PMC and PEC systems can be understood as follows. Because of the boundary conditions at the interface between the DNG slab and the PMC mirror, the PMC system is equivalent to the symmetric system with two *parallel* emitting dipoles on both sides of the slab and one absorbing nanocylinder at the center considered in

[6], where majority of the energy from the two emitting dipoles would be focused onto the nanocylinder. Likewise, in the PMC system [Fig. 2(a)], almost all the energy radiated by the emitting dipole towards the slab would flow to the absorbing nanocylinder in the region of $0.1 < \varepsilon'' < 0.5$.

In contrast, as dictated by the boundary conditions at the interface between the DNG slab and the PEC mirror, the PEC system is equivalent to the symmetric system with two *antiparallel* emitting dipoles on both sides of the slab and one absorbing nanocylinder at the center. In this case the nanocylinder is no longer a focal point and that is why it cannot absorb effectively.

Surprisingly, Fig. 2 shows that in both the PMC and PEC systems, more than half of the energy radiated by the single dipole source is absorbed by the whole system (the nanoparticle plus the slab) in the entire region of the imaginary part of the cylinder permittivity considered. Therefore, both systems can be considered as perfect absorbers, a fact which was overlooked before [6]. In fact, the power radiated by the dipole source is almost completely absorbed by the slab and nanoparticle together in the PEC system, which therefore may be called a super-absorber. More detailed analysis of this possibility will be presented in section 3.

*2.2 Source dipole perpendicular to the metamaterial slab*

Now let us consider the case of the radiating electric dipole $p_0$ being perpendicular to the surface of the metamaterial slab. In this case, the dipole nanowire has a current density given by

$$j_x = -i\omega p_0 \delta(y)\delta(x+l). \qquad (3)$$

That is, we are again considering a 2D geometry with only nonzero $H_z$ component of the magnetic field. The power absorbed by the central nanoparticle ($W_c/W_0$), the slab ($W_s/W_0$) and by the whole system (i.e., the nanoparticle plus the slab) ($W_{c+s}/W_0$) as well as the power radiated from the source dipole ($W_p/W_0$) for the systems with either a PEC or PMC mirror, are displayed in Fig. 3.

Like the system of the source dipole moment parallel to the slab surface with the PMC mirror, more than half of the power emitted by the source dipole was absorbed by the absorbing nanoparticle in the PEC system in the $\varepsilon_c''$ range of 0.06~0.6[see Fig. 3(a)]. Therefore, the PEC system can serve a perfect nanoabsorber for the source dipole pointing towards the slab, a case which was not considered in [6]. As mentioned before, the system can also be used as a perfect nanodetector if the nanocylinder is a detecting element.

In contrast, the power absorbed by the nanoparticle in the PMC system is much smaller than half of the energy radiated by a single source in the entire region of the imaginary part of the cylinder permittivity considered here. Consequently, the PMC system does not exhibit perfect nanoabsorption or nanodetection behavior.

These contrasting behaviors of the PEC and PMC systems can be understood again by making analogous to the symmetric system of two point sources and one absorbing nanoparticle. Due to the boundary conditions at the interface between the DNG slab and the PEC mirror, the PEC system is equivalent to the symmetric system with two *parallel* emitting dipoles on both sides of the slab and one absorbing nanocylinder at the center. In contrast, as dictated by the boundary conditions at the interface between the DNG slab and the PMC mirror, the PMC system is equivalent to the antisymmetric system with two *antiparallel* emitting dipoles on both sides of the slab and one absorbing nanocylinder at the center. These two cases with the dipole moments perpendicular to the DNG slab were not investigated in [6,9-10]. Nevertheless, a straightforward extension of the previous analytical work in [6,9-10] show that the system of two *parallel* emitting dipoles on both sides of the slab and one absorbing dipole (or nanoparticle) at the center of slab would satisfy the Maxwell equations, i.e., all the energy from the two sources would go to the sink. On the other hand, the system of two *antiparallel* dipole sources and one absorbing dipole sink is not a solution of the Maxwell equations. In other words, there is no focusing at the sink in this case. Therefore, in the PEC system, majority of the energy radiated by the emitting dipole towards the slab would flow to the absorbing nanocylinder, while in the PMC system, the nanocylinder is no longer a focal point and cannot absorb energy efficiently.

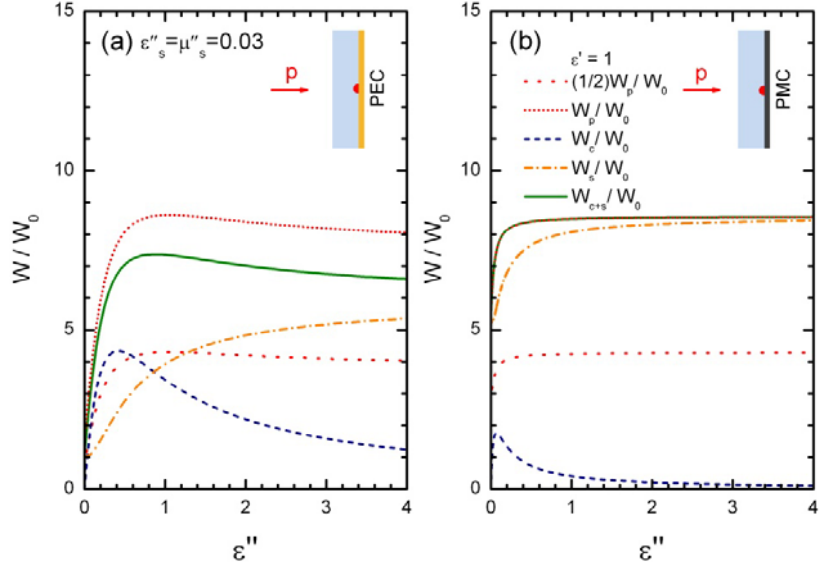

Fig. 3. Powers absorbed by the nanoparticle ($W_c/W_0$), the slab ($W_s/W_0$) and the system (nanoparticle plus slab) ($W_{c+s}/W_0$) normalized to the radiating power of one dipole in free space ($W_0$) as a function of the imaginary part of the cylinder permitivity for the system with either a PMC mirror (a) or a PEC mirror (b) at the back of the metamaterial slab. The radiating electric dipole is perpendicular to the surface of the DNG slab. The total ($W_p/W_0$) and half [($W_p/2)/W_0$] normalized radiating powers of the dipole are also plotted for comparison. The radius of nanoparticle ($r_c$) is 50 nm, and its optical parameters are $\varepsilon_c = 1 + i\varepsilon''$ and $\mu_c = 1$, $\varepsilon_s = \mu_s = -1 + 0.03i$, $\lambda = 3$ μm, $d = 250$ nm and $l = 250$ nm.

Interestingly, Fig. 3 shows that in both the PEC and PMC systems, more than half of the energy radiated by the single source dipole is absorbed by the system (the nanoparticle plus the slab) in the entire region of the imaginary part of the cylinder permittivity considered. As a result, both systems can be considered perfect absorbers. In fact, almost all the power radiated by the dipole source is absorbed by the slab and nanoparticle together in the PMC system, which therefore may be called a super-absorber (for details see section 3).

An examination of Fig. 2(a) and Fig. 3(a) reveals some differences between the PMC system with the parallel electric dipole and the PEC system with the perpendicular dipole. First, the range of the imaginary part of the cylinder dielectric constant that the power absorbed by the nanocylinder is larger than half of the power emitted by the radiating dipole, is larger in the latter system than in the former system. Second, in the common region of the strong absorption, the radiating power of the source dipole is smaller in the PEC system than in the PMC system. This means that less energy output from the emitting dipole in the PEC system than in the PMC system, and hence the PEC system would be more sensitive and energy-saving than the PMC system. Furthermore, in practice, it is much easier to make a PEC mirror than a PMC mirror. Therefore, if these systems are to be used as coherent nanoabsorption or nanodetection devices for converging light to a sample at the nanocylinder, the PEC system would be more useful.

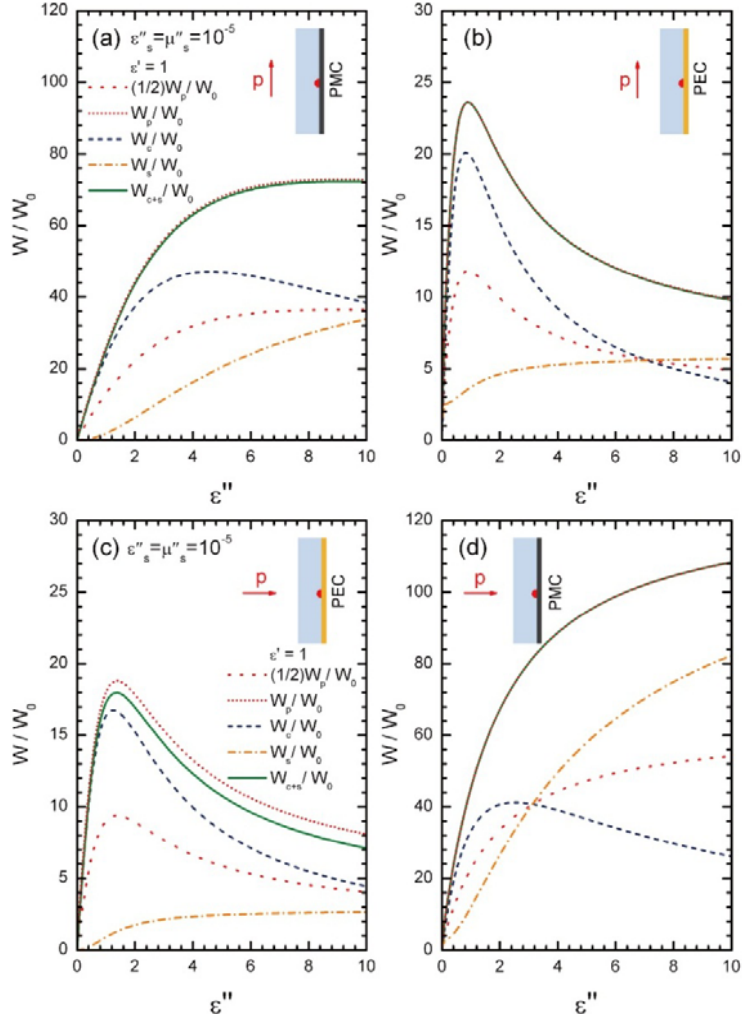

Fig. 4. Powers absorbed by the nanoparticle ($W_c/W_0$), the slab ($W_s/W_0$) and the system (nanoparticle plus slab) ($W_{c+s}/W_0$) normalized to the radiating power of one dipole in free space ($W_0$) as a function of the imaginary part of the cylinder permittivity for the system with either a PMC mirror or a PEC mirror at the back of the metamaterial slab. The radiating electric dipole is parallel (a, b) or perpendicular (c, d) to the surface of the DNG slab. The total ($W_P/W_0$) and half [($W_P/2)/W_0$] normalized radiating powers of the dipole are also plotted for comparison. The radius of nanoparticle ($r_c$) is 50 nm, and its optical parameters are $\varepsilon_c = 1 + i\varepsilon''$ and $\mu_c = 1$, $\varepsilon_s = \mu_s = -1 + 10^{-5}i$, $\lambda = 3$ μm, $d = 250$ nm and $l = 250$ nm.

## 2.3 Effect of the losses of the metamaterial slab

Surprisingly, Figs. 2 and 3 show that in a wide range of the imaginary part $\varepsilon_c''$ of the permittivity of the nanocylinder, the energy absorbed by the slab is substantially larger than the energy absorbed by the nanocylinder in all the systems considered, even though the slab is with small losses ($\varepsilon_s''=0.03$). Therefore, to enhance the focusing property of the PMC system with the parallel emitting dipole and also the PEC system with the perpendicular dipole, one should further reduce the losses of the slab. We therefore perform a series of the FEM simulations using the $\varepsilon_s^{''}$ ranging from $10^{-5}$ to 0.03.

The power absorbed by the central nanoparticle ($W_c/W_0$), the slab ($W_s/W_0$) and the whole system (i.e., the nanoparticle plus the slab) ($W_{c+s}/W_0$) as well as the power radiated from the source dipole ($W_p/W_0$) for the systems with negligible slab losses ($\varepsilon_s^{''} = 10^{-5}$) for both the electric dipole parallel to and perpendicular to the slab surface, are displayed in Fig. 4. Here, first of all, one can see that due to the increased quality factor of the surface plasmon waves, the total radiated energy is substantially increased in comparison with the previous case of greater losses (see Fig. 2 and Fig. 3). Figure 4 shows clearly that this enhanced power is absorbed by the nanocylinder effectively for any dipole orientations and it is larger than half of the radiated power in a wide range of the $\varepsilon_c^{''}$ from 0.0 to ~10.0 as we expected indeed. Also, in this $\varepsilon_c^{''}$ range, the power absorbed by the slab is now substantially smaller than that by the nanocylinder.

Rather unexpectedly, Fig. 4(b) [Fig. 4(d)] indicates that in the PEC (PMC) system with the dipole parallel (perpendicular) to the slab surface, the power absorbed by the nanocylinder is also larger than half of the power radiated by the emitting dipole in the $\varepsilon_c^{''}$ range of 0.01~7.0 (0.01~3.1). This means that the PEC (PMC) system with the dipole parallel (perpendicular) to the slab surface could also be used as a perfect nanoabsorbers.

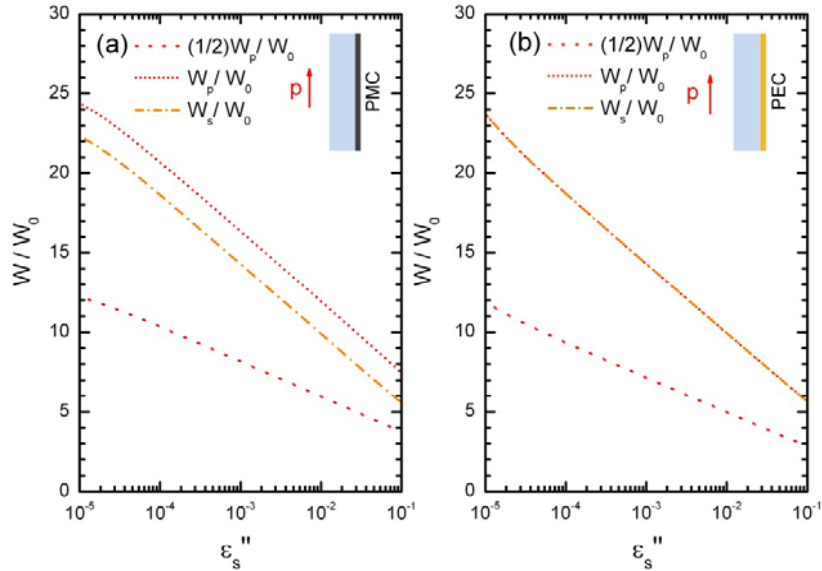

Fig. 5. Power absorbed by the slab ($W_s/W_0$) as well as the total ($W_p/W_0$) and half [$(W_p/2)/W_0$] radiating powers of the dipole normalized to the radiating power of one dipole in free space ($W_0$) as a function of the imaginary part of the slab permittivity for the system with either a PMC mirror (a) or a PEC mirror (b) at the back of the metamaterial slab. The electric dipole is parallel to the surface of the DNG slab.

Let us stress again that the power radiated by the radiating electric dipole in all the systems considered is significantly larger than the radiation power of a single dipole in free space (see Figs. 2-4). This is due to the presence of the environment such as the absorbing nanocylinder and the lossy slab, the well known Purcell effect [9]. This enhancement of the radiation of the electric dipole ranges from a few folds to slightly more than 10 in the systems with the DNG slab with small losses ($\varepsilon_s^{''}=0.03$) (see Figs. 2-3), and also in the PEC systems with the DNG slab of negligible losses ($\varepsilon_s^{''} = 10^{-5}$) [see Fig. 5(b) and Fig. 6(a)]. However, in the PEC systems consisting of the DNG slab with negligible losses, the radiation power of the emitting dipole and hence the absorption power of the nanocylinder decreases steadily as the imaginary part of the nanocylinder dielectric constant increases [Figs. 4(b) and 4(c)]. Interestingly, the enhancement of the radiation power of the single dipole due to the Purcell effect could reach ~70 in the PMC system with the emitting dipole parallel to the slab surface [Fig. 4(a)] and also reach more than 100 in the PMC system with the perpendicular emitting dipole [Fig. 4(d)].

## 3. Metamaterial slab as a super-absorber

As mentioned before, more than half of the energy radiated by the single dipole source is absorbed by the slab with even small losses in the entire region of the imaginary part of the cylinder permittivity considered (see Fig. 2 and Fig. 3). Furthermore, in the cases of the emitting dipole parallel to the slab with the PEC mirror and the emitting dipole perpendicular to the slab with the PMC mirror, the absorption by the slab dominates over the absorption by the nanoparticle. This is mainly because there is no focusing on the nanoparticle in these two cases. To further examine this important role in the absorption of the dipole radiation played by the metamaterial slab, we have performed the analytical and the FEM simulations for all the four systems without the nanoparticle. The results for the emitting dipole parallel to and also perpendicular to the metamaterial slab are shown in Fig. 5 and Fig. 6, respectively.

Figures 5(a) and 6(a) show that around 75-90 % power radiated by the single dipole source is absorbed by the metamaterial slab in the cases of the emitting dipole parallel to the slab surface with the PMC mirror and also of the emitting dipole perpendicular to the slab surface with the PEC mirror. Therefore, both systems can be called perfect absorbers, in the terms of the terminologies in [6]. Remarkably, Fig. 5(b) [Fig. 6(b)] indicates that the PEC (PMC) system absorbs all the energy radiated by the single source with the dipole parallel (perpendicular) to the slab surface. Therefore, both these systems may be considered as a super-absorber.

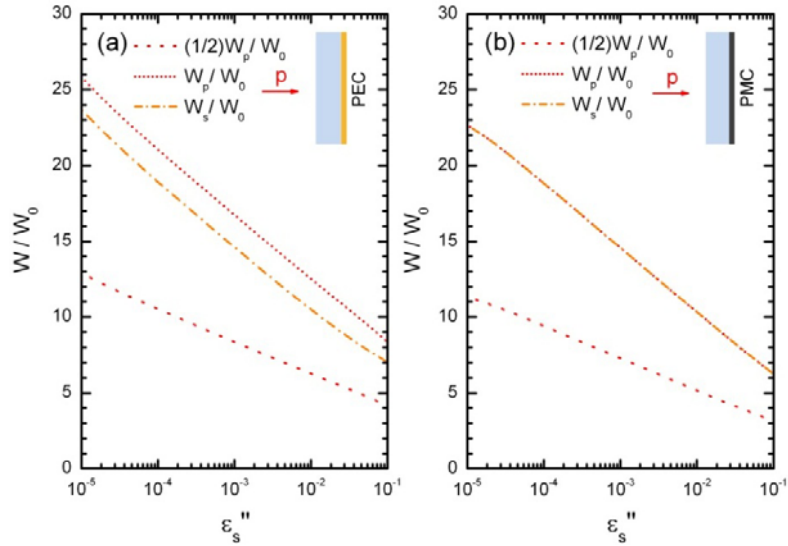

Fig. 6. Power absorbed by the slab ($W_s/W_0$) as well as the total ($W_p/W_0$) and half ($W_p/2W_0$) radiating powers of the dipole normalized to the radiating power of one dipole in free space ($W_0$) as a function of the imaginary part of the slab permittivity for the system with either a PEC mirror (a) or a PMC mirror (b) at the back of the metamaterial slab. The electric dipole is perpendicular to the surface of DNG slab.

Interestingly, Fig. 5 and Fig. 6 indicate that the absorption power of the DNG slab is almost the same in all the four systems we consider here. However, the dipole radiation power in the PMC system with the dipole parallel to the slab surface [Fig. 5(a)] and also in the PEC system with the dipole perpendicular to the slab surface [Fig. 6(a)] is larger by about $2W_0$ (twice the radiation power of one dipole in free space) than that in the PEC system with the dipole parallel to the slab surface [Fig. 5(b)] and also in the PMC system with the dipole perpendicular to the slab surface [Fig. 6(b)].

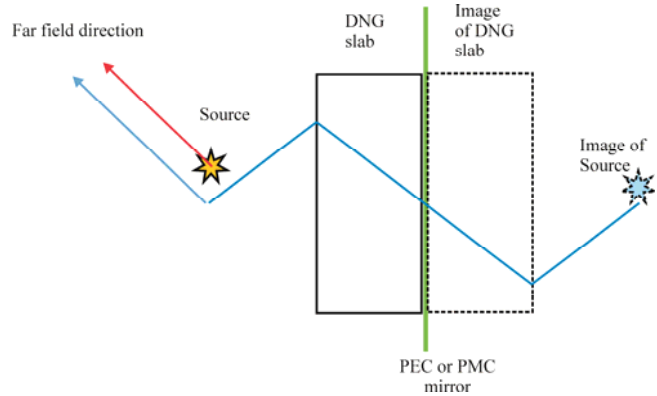

Fig. 7. Illustration of the zero phase difference between the source and its image for any far field directions.

This remarkable behavior of the DNG slab without the nanoparticle has a simple physical explanation. Due to the perfect nature of the PEC or PMC mirror, the fields in the system under consideration are equal to the fields in a system without the mirror but with the mirror slab and source, as illustrated in Fig. 7. It is easily seen from this figure that due to the negative refractive index of the DNG slab, for any far field direction, the positive phase difference ($+2\Delta$) between the source and its image is fully compensated by the negative phase shift ($-2\Delta$) due to the propagation inside the DNG slab. This means that for the far field radiation, the source and its image can be considered as situated at the same point in space. Now in the cases of Figs. 5(a) and 6(a) where the dipole and its image have the same orientation, the radiation into the half-space under consideration will be equal to $4W_0/2=2W_0$. The rest of the radiated power will be in the near field form and hence will be absorbed by the DNG slab. In contrast, for the cases of Figs. 5(b) and 6(b) where the dipole and its image have the opposite orientations, the radiation into the half-space under consideration will be zero. Therefore, all the emitted energy will be absorbed by the DNG. This is what exactly can be observed in Figs. 5- 6.

## 4. Closing remarks and summary

In this paper, we have investigated the possibility to build effective super-absorbers and perfect nanodetectorson the base of the DNG metamaterials. Both analytical dipole model analyses and numerical simulations have shown that an absorbing nanoparticle properly placed inside a DNG metamaterial slab can absorb up to 100% radiated energy of a single divergent source placed outside of the slab. This allows us to put forward the concept of the single side perfect nanoabsorber (PNA) for diverging beams. Furthermore, in contrary to CPA with a usual dielectric cavity, the PNA device proposed here is robust to small perturbations in the properties of the absorbing nanoparticle and the metamaterial. The generalization of this approach to the 3D systems is straightforward. Furthermore, our concept can be generalized to more complicated geometries such as wedges and spheres.

We would like to stress again that our systems allow focusing the radiation in nanoscale regions and can be applied in the optical nanodevices for different purposes. For example, one may use a single side PNA to arrange readout of the results of quantum computation which are based on single photon qubits.

Finally, we have also discovered that even a simple DNG slab with a PEC or PMC mirror but without any absorbing nanoparticle can be used as a highly efficient absorber for both plane and divergent beams.

**Acknowledgments**

G. Y. Guo, S. Sun and W.-J. Zheng would like to thank the National Science Council and National Center for Theoretical Sciences of Taiwan as well as Center for Quantum Science and Engineering, National Taiwan University (CQSE-10R1004021) for financial supports. V. Klimov would like to express his gratitude to the Russian Foundation for Basic Research (grants ## 11-02-91065, 11-02-92002, 11-02-01272, 12-02-90014), the Russian Quantum Center and Skolkovo foundation for financial supports.